\documentstyle[preprint,aps,graphicx]{revtex}

\makeatletter
\def\hlinewd#1{\noalign{\ifnum0=`}\fi
\hrule \@height #1 \futurelet \reserved@a\@xhline}
\def\hwhiteline{\noalign
{\ifnum0=`}\fi\hrule
\@height 0pt\vskip 1.0ex\futurelet \reserved@a\@xhline}
\makeatother
\def\gray{\special{ps: 0.40 setgray}}
\def\black{\special{ps: 0.0 setgray}}
\def\red{
}
\newcommand{\mydraft}{
\newcount\timecount
\newcount\hours \newcount\minutes  \newcount\temp \newcount\pmhours

\hours = \time
\divide\hours by 60
\temp = \hours
\multiply\temp by 60
\minutes = \time
\advance\minutes by -\temp
\def\hour{\the\hours}
\def\minute{\ifnum\minutes<10 0\the\minutes
       \else\the\minutes\fi}
\def\clock{
\ifnum\hours=0 12:\minute\ AM
\else\ifnum\hours<12 \hour:\minute\ AM
\else\ifnum\hours=12 12:\minute\ PM
       \else\ifnum\hours>12
        \pmhours=\hours
        \advance\pmhours by -12
        \the\pmhours:\minute\ PM
        \fi
       \fi
\fi
\fi
}
\def\fullclock{\hour:\minute}
\begin{centering}
\gray
\font\Hugett  =cmtt12 scaled\magstep4
\hbox{\Hugett Draft:\today,\clock}
\black
\end{centering}
\vskip -1.7cm
$\phantom{a}$
} 

\def\beq#1{\begin{equation} \label{#1}}
\def\eeq{\end{equation}}

\newskip\humongous \humongous=0pt plus 1000pt minus 1000pt

\newif\ifdtup

\def\TT{\hbox{\small $\bar{\bf 3}{\bf 3}$}}
\def\SS{\hbox{\small $\bar{\bf 6}{\bf 6}$}}

\def\3s{\hbox{\small $\bar{\bf s}\bar{\bf 3}$}}
\def\6s{\hbox{\small $\bar{\bf s}{\bf 6}$}}

\def\mycomm#1{\hfill\break\strut\kern-3em{\red\tt ====> #1\black}\hfill\break}
\def\lowstrut{\vrule height 0ex depth 0.9ex width 0pt } 
\def\lowstrutA{\vrule height 0ex depth 1.1ex width 0pt } 
\def\upstrut{\vrule height 2.2ex depth 0.0ex width 0pt } 
\def\Tcc{\hbox{$\,{T^{\pm}\kern-2.0ex\lowstrut}_{\bar c c}\,$}}
\def\Tbb{\hbox{$\,{T^{\pm}\kern-2.0ex\lowstrutA}_{\bar b b}\,$}}
\begin{document}
{\tighten
    \preprint {\vbox{
     \hbox{$\phantom{aaa}$}
     \vskip-0.5cm
\hbox{TAUP 2869/07}
\hbox{WIS/03/08-FEB-DPP}
\hbox{ANL-HEP-PR-08-7}
}}

\title{Possibility of Exotic States in the Upsilon system}
\author{Marek Karliner\,$^{a}$\thanks{e-mail: \tt marek@proton.tau.ac.il}
\\
and
\\
Harry J. Lipkin\,$^{a,b}$\thanks{e-mail: \tt
ftlipkin@weizmann.ac.il} }
\address{ \vbox{\vskip 0.truecm}
$^a\;$School of Physics and Astronomy \\
Raymond and Beverly Sackler Faculty of Exact Sciences \\
Tel Aviv University, Tel Aviv, Israel\\
\vbox{\vskip 0.0truecm}
$^b\;$Department of Particle Physics \\
Weizmann Institute of Science, Rehovot 76100, Israel \\
and\\
High Energy Physics Division, Argonne National Laboratory \\
Argonne, IL 60439-4815, USA\\
}
\maketitle
\thispagestyle{empty}

\begin{abstract}%
\strut\vskip-1.0cm
Recent data from Belle show 
unusually large partial widths
$\Upsilon(5S) \to \Upsilon(1S)\, \pi^+\pi^-$
and
$\Upsilon(5S) \to \Upsilon(2S)\, \pi^+\pi^-$.
The $Z(4430)$ narrow resonance also reported by Belle 
in $\psi^\prime \pi^+$ spectrum
has the
properties expected of a $\bar c c u \bar d$ charged isovector tetraquark
\Tcc\,.
The analogous state \Tbb\ in the bottom sector
might mediate
anomalously large cascade decays in the Upsilon system,
$\Upsilon(mS) \to \Tbb \pi^{\mp}\to \Upsilon(nS)\, \pi^+ \pi^-$,
with a tetraquark-pion intermediate state.
We suggest looking  for the $\bar b b u \bar d$ 
tetraquark in these decays as peaks in the invariant mass of 
$\Upsilon(1S)\,pi$ or $\Upsilon(2S)\,\pi$ systems.
The $\bar b b u \bar s$ 
tetraquark can appear in the observed  decays 
$\Upsilon(5S) \to \Upsilon(1S) \, K^+K^-$ as a peak in the invariant mass of 
$\Upsilon(1S)\, K$ system.
We review the model showing that these tetraquarks are below the two heavy meson
threshold, but respectively above the $\,\Upsilon\,\pi\pi$ and 
$\,\Upsilon\,K \bar K $ thresholds. 

\end{abstract}

\vfill\eject

\section{Introduction}

Conclusive evidence for the first unambiguously multiquark states would be
obtained by finding isovector states or strange states containing heavy quark 
pairs. The presence of possible tetraquark states containing heavy quarks has
been extensively discussed. Simple calculations\cite{newpentel} indicate that 
such tetraquarks containing heavy quark-antiquark pairs might be observed as
pion-charmonium or pion-bottomonium resonances. 
Such resonances have been observed  above the
threshold for decays into two heavy mesons. For the bottomonium system
our calculations predict tetraquarks with masses below the threshold for decay into
two heavy mesons but above the threshold for decay into heavy bottomonium plus a
pion. These can be observed as bottomonium pion or bottomonium kaon resonances
in the cascade decays from the higher $\Upsilon(nS)$ resonances.

The search for intermediate isovector heavy-quarkonium states has been going on
for a long time without any positive results. Previous work suggesting possible
isovector exotic states as intermediate states in the $\pi\pi$ cascade decays
of heavy quarkonium resonances began with  $\psi' \to \pi \pi J/\psi$ and was
later extended to the $\Upsilon$  system \cite{voloshin,liptuan}. Unfortunately
none of these suggestions were confirmed by later
experiments\cite{zou,guo,CLEO}. No $\pi J/\psi$ nor $\pi \Upsilon$ resonances
were found.  No theoretical model predicted that tetraquark states would exist
close enough to the two-heavy-meson threshold that so they would not be
impossibly broad.

New information has recently become available suggesting that these exotic
states might now be observable: 

\begin{enumerate}
\item New data for the $\Upsilon (5S)$ decays\cite{UpsilonDecays}
have a serious $\pi\pi$ problem.
\item A new theoretical model\cite{newpentel} predicts isovector tetraquark
states around the $B\bar B$ threshold.
\end{enumerate}

     That such tetraquarks should exist for bottomonium and not for
charmonium is predicted in this model which has unconventional color
couplings and color-space correlations. The two quarks are
in a color sextet, the two antiquarks are in a color antisextet, and the
color-space correlations require the mean quark-antiquark distance
to be considerably smaller than the mean quark-quark and
antiquark-antiquark distances.

\section{New experimental data suggesting the existence of tetraquarks}

\subsection{Peculiarities in $\Upsilon(\MakeLowercase{n}S)$ production}

The Belle Collaboration has recently reported\cite{UpsilonDecays} 
anomalously large partial widths in 
$\Upsilon(1S) \,\pi^+ \pi^-$
and
$\Upsilon(2S) \,\pi^+ \pi^-$
production at the $\Upsilon(5S)$, 
more than two orders of magnitude larger than the corresponding 
partial widths for $\Upsilon(4S)$, $\Upsilon(3S)$ or $\Upsilon(2S)$
decays.

We suggest that the large partial widths of
these channels might be due to their production by decays via an 
intermediate 
$\Tbb\pi^{\mp}$ state, where \Tbb\ denotes an isovector charged
 tetraquark $\bar b b u \bar d$ or $\bar b b \bar u d$,
 
\beq{upstet}
\Upsilon(nS)\ \rightarrow \ \pi^{\mp} \Tbb \ \rightarrow \ 
\Upsilon(mS)\,\pi^-\pi^+ 
\eeq
\subsection{The $Z(4430)$ resonance} 

In the summer of 2007 Belle reported \cite{Z4430} a narrow 
resonance-like structure $Z(4430)$ in the $\psi^\prime \pi^{\pm}$ 
invariant mass with mass and width
$M=4433 \pm 4\hbox{(stat)}\pm 2\hbox{(syst)}$ MeV and
\hbox{$\Gamma= 45{\,}^{+18}_{{-}13}\,\hbox{(stat)}
\, {}^{+30}_{{-}13}\,\hbox{(syst)}$ MeV.} 
The $Z(4430)$ has not been seen in the $J/\psi \pi^\pm$ channel.
This report is awaiting
confirmation. If it is confirmed, it can be 
a $\bar c c u \bar d$ tetraquark,
since it is an isovector and carries hidden charm.
Most calculations predict that such states are above the masses of two 
separated heavy quark mesons as well as $Q \bar Q$ and
$u \bar d$ mesons. The $cu\bar c \bar d$ and  $bu\bar b\bar d$ states can 
therefore decay into states like $D \bar D$ and $B \bar B$ as well as
$J/\psi \pi$ and $\Upsilon \pi$ with large widths. The narrow width
of the $Z(4430)$ 
and the lack of the $J/\psi \pi^\pm$ decay channel are therefore quite puzzling.
However a mechanism
\cite{mix} has been proposed in which the two color eigenstates can be mixed in
such a way that the otherwise dominant $D \bar D$ and $B \bar B$ are suppressed.   
If confirmed, the existence of $Z(4430)$ would also make it very likely
that there is an analogous state in the bottom system.

$Z(4430)$ is approximately 700 MeV above the $D D$ threshold. But the
new Belle data suggest the existence of bottom tetraquarks 
not far from the $B B$ threshold, i.e.
at much lower mass than the naive bottom analogue of $Z(4430)$.

\section{Models for the 
\boldmath $T^+ \lowercase{(b u \bar b \bar d)} \unboldmath $ and  
\boldmath $T_s^+ \lowercase{(b u \bar b \bar s)} \unboldmath$ tetraquarks}

Resonant states containing heavy quark $Q\bar Q$ 
components are expected to decay with large
widths into heavy quarkonium $Q\bar Q$ states with one or two additional pions; 
e.g $J/\psi \pi$, $J/\psi \pi\pi$, $\Upsilon \pi$, and $\Upsilon \pi\pi $ 
if phase space is available. 

One example is the case of the
$Qu\bar Q \bar d$ tetraquarks whose masses have been
shown in a number of cases to be comparable to the masses of two separated
heavy-quark mesons\cite{newpentel}. 

According to \cite{newpentel}, the $b u \bar b \bar d$ tetraquark can 
have a mass
below the mass of two $B$ mesons and above the mass of the $\Upsilon$.
It can decay into $\pi \Upsilon$. The $\Upsilon (2S)$ and $\Upsilon(3S)$ are 
both
below the $B \bar B$ threshold and could decay into an $I=1$ tetraquark and a
pion. The $I=1$ tetraquark could then decay into the $\Upsilon(1S)$ and a pion
and might decay into an $\Upsilon(2S)$ and a pion. 

The partial widths of $\Upsilon(5S)$ into $\Upsilon(nS)\,\pi^+\pi^-$
are reported \cite{UpsilonDecays} to be more than two orders of magnitude
larger than the corresponding partial widths for 
$\Upsilon(4S)$, $\Upsilon(3S)$ or $\Upsilon(2S)$ decays.
{\em We suggest looking for the $\bar b b u \bar d$ 
tetraquark in these decays as peaks in the invariant mass of 
$\,\pi\Upsilon(nS)$ systems.}

The $b u \bar b \bar d$ configuration was treated in ref. \cite{newpentel} in a
harmonic oscillator model with the Nambu interaction. There are two color
couplings for a color singlet tetraquark state of two quark antiquark pairs.
The two quarks can be coupled to either a color antitriplet or a color sextet,
with the two antiquarks coupled to conjugate representations. The
triplet-antitriplet tetraquark is  denoted by $\TT$, and a sextet-antisextet
tetraquark denoted by $\SS$. 

The result for the ratio of
the ground state energy of the $b u \bar b \bar u$ tetraquark to the ground 
state energy of the separated $B \bar B$ two-meson system was found to be 
\beq{EDDbuburat}
{{E_g(\TT\,\bar b u b \bar u)}\over{\upstrut E_g(2M(\bar b u) )}}
=1.042\,; 
\qquad\qquad\qquad
{{E_g(\SS\,\bar b u b \bar u)}\over{\upstrut E_g(2M(\bar b u) )}}
= 0.891
\end{equation}
where $\TT$ and $\SS$ denote respectively the triplet-antitriplet and
sextet-antisextet
tetraquark states.  

The mass of the $b u \bar b \bar d$ tetraquark with the color sextet-antisextet
coupling is found to be well below the two-meson $B\bar B$ threshold in this
approximation. Such a low-mass threshold may also be found in a more exact
calculation including spin effects. That they might be found in the experimental
spectrum must be taken seriously.

The Belle Collaboration has also reported\cite{UpsilonDecays}  $K^+K^-
\Upsilon(1S)$  production at the $\Upsilon(5S)$. Such states can be produced by
decays via an  intermediate  $T_s^{\pm}{K\mp}$ state where $T_s^{\pm}$ denotes
a strange $\bar b b u \bar s$ tetraquark

\beq{upsstrtet}
\Upsilon(nS)\rightarrow K^{\mp} T_s^{\pm} \rightarrow \Upsilon(mS)\,
K^-K^+
\eeq
{\em We suggest looking for the $\bar b b u \bar s$ 
tetraquark in these decays as peaks in the invariant mass of 
$K\Upsilon(1S)$ or $K\Upsilon(2S)$ systems.}

The $b s \bar u \bar b$ tetraquark configuration was not treated in 
ref. \cite{newpentel}. We apply the same analysis here by simply changing the
quark labels and masses, obtaining 
\beq{EDDbubsrat}
{{E_g(\TT\,\bar b u b \bar s)}\over{\upstrut E_g[ (\bar b u)\,(b \bar s) ]}}
=1.057\,; 
\qquad\qquad\qquad
{{E_g(\SS\,\bar b u b \bar s)}\over {\upstrut E_g[ (\bar b u)\,(b \bar s) ]}}
= 0.914
\end{equation}
Details of the calculation are given in the Appendix.

The mass of the $b u \bar b \bar s$ tetraquark with the color sextet-antisextet
coupling is found to be well below the two-meson $B_s\bar B$ threshold in this
approximation. 
\section{Conclusion}

Conclusive evidence for the first unambiguous multiquark states would be
obtained by finding isovector states or strange states containing heavy quark 
pairs. Model calculations suggest that such states should exist with masses
below the  $B\bar B$ threshold but above the $\pi \Upsilon(nS)$ or
$K\Upsilon(nS)$  thresholds. Searches are suggested for these states by
measuring the invariant masses of $\Upsilon(nS)\,\pi$ and 
$\Upsilon(nS)\,K$ pairs
produced in the cascade decays of a higher $\Upsilon(nS)$ resonance to lower
$\Upsilon(nS)$ states by pion pair or kaon pair emission. Such transitions are
suggested by  present data indicating  anomalous production of 
$\Upsilon(nS)\,\pi\pi$ states in  $\Upsilon(5S)$  decays. 

\section*{Appendix}

The ground state energies of the 
$b s \bar u \bar b$
states are given by the sum of the ground
state energies of the contributing harmonic oscillators. The ground state
energy of an oscillator with co-ordinate $r$, mass $M$
and potential $\,Vr^2\,\,$  is

    \beq{gs}
E_g(osc) = {{3}\over{2}}\cdot\hbar \omega =
{{3}\over{2}}\cdot\hbar\sqrt{{{V}\over{M}}}
\end{equation}

For the two-meson
$(\bar b u;\,b \bar s$)
system, which has two separated harmonic oscillators with reduced masses
denoted by $M(bu)$ and
$M(bs)$ and potentials
$V_o/2$ from ref. \cite{newpentel},

\beq{E2Msucd} 
E_g[ (\bar b u)\,(b \bar s) ]=
{{3}\over{2}}\cdot\hbar
\left(\sqrt{{{V_o}\over{2M(bu)}}}+
\sqrt{{{V_o}\over{2M(bs)}}}\,\,\right)
\end{equation}

The ground state energies for the $\TT$ and $\SS$ systems are obtained by
substituting the potentials from ref. \cite{newpentel}  into the
expression (\ref{gs}) along with  the reduced masses

    \beq{redmassc}
M(bs)={{m_b m_s}\over{m_b +m_s}}; ~ ~ ~  M(bu)={{m_b m_u}\over{m_b +m_u}};
    ~ ~ ~ M\{(bs)(bu)\}
    = {{(m_b +m_s)\cdot(m_b +m_u) }\over{m_b +m_s + m_b +m_u}}
\end{equation}

\beq{EDD}
E_g(\TT) = {{3}\over{2}}\cdot\hbar
\left(\sqrt{{{V_o}\over{2M\{(bs)(bu)\}}}}+
\sqrt{{{3V_o}\over{8M(bs)}}}+
\sqrt{{{3V_o}\over{8M(bu)}}}\,\,\,\right)
\end{equation}
\beq{Etetra1}
E_g(\SS) = {{3}\over{2}}\cdot\hbar
\left(\sqrt{{{5V_o}\over{4M\{(us)(bu)\}}}}+
\sqrt{{{3V_o}\over{16 M(bs)}}}+
\sqrt{{{3V_o}\over{16 M(bu)}}}\,\,\,\right)
\end{equation}
The ratios of these energies to the energy of the two meson state are then

\beq{EDDsucdrat0}
\begin{array}{l}
\displaystyle
{{E_g(\TT\,\bar b u b \bar s)}\over
\upstrut E_g[ (\bar b u)\,(b \bar s) ] }=
\\
\hfill\\
\displaystyle
\sqrt{3\over4}
+ {{\sqrt{(m_b +m_s + m_b +m_u)m_sm_bm_u}}\over {m_b (m_u-m_s)}}
\cdot
\displaystyle
\left[\sqrt{{{m_u}\over{m_b +m_u}}}-
\sqrt{{{m_s}\over{m_b +m_s}}}
\,\,\,\right]= 1.057
\end{array}
\end{equation}

\beq{Etetra(cusurat)}
\begin{array}{c}
\displaystyle
{{E_g(\SS\,\bar b u b \bar s)}\over
\upstrut E_g[ (\bar b u)\,(b \bar s) ] }=
\hfill\\
\\
\displaystyle
\strut\kern-1em
= \sqrt{{{3}\over{8}}}+
\sqrt{5\over 2}\cdot
{{\sqrt{(m_b +m_s + m_b +m_u)m_sm_bm_u}}\over {m_b (m_u-m_s)}}
\cdot
\displaystyle
\left[\sqrt{{{m_u}\over{m_b +m_u}}}-
\sqrt{{{m_s}\over{m_b +m_s}}}
\,\,\,\right]
=0.914
\end{array}
\end{equation}
where we have  substituted the  values of
the constituent quark masses obtained by
fitting the ground state meson and baryon spectra\cite{NewPenta}.

\beq{massval} m_b =  360\ {\rm MeV} ; ~ ~ ~
m_s=  540\ {\rm MeV}  ; ~ ~ ~ m_u= 1710 \ {\rm MeV}; ~ ~ ~  m_b =  5050
\ {\rm MeV}
\end{equation}

 \section*{Acknowledgements}
The research of M.K. was supported in part by a grant from the
Israel Science Foundation administered by the Israel
Academy of Sciences and Humanities.
We thank 
 Bill Dunwoodie,
Inga Karliner,
Steve Olesen,
Jon Thaler 
and  Misha Voloshin
for useful discussions.
\vfill\eject

%
\catcode`\@=11 
\def\references{
\ifpreprintsty \vskip 10ex
%
\hbox to\hsize{\hss \large \refname \hss }\else
\vskip 24pt \hrule width\hsize \relax \vskip 1.6cm \fi \list
{\@biblabel {\arabic {enumiv}}}
{\labelwidth \WidestRefLabelThusFar \labelsep 4pt \leftmargin \labelwidth
\advance \leftmargin \labelsep \ifdim \baselinestretch pt>1 pt
\parsep 4pt\relax \else \parsep 0pt\relax \fi \itemsep \parsep \usecounter
{enumiv}\let \p@enumiv \@empty \def \theenumiv {\arabic {enumiv}}}
\let \newblock \relax \sloppy
    \clubpenalty 4000\widowpenalty 4000 \sfcode `\.=1000\relax \ifpreprintsty
\else \small \fi}
\catcode`\@=12 

\end{document}